# Bright solitons from defocusing nonlinearities

Olga V. Borovkova,[1] Yaroslav V. Kartashov,[1] Lluis Torner,[1] and Boris A. Malomed[2,1]

[1]ICFO-Institut de Ciencies Fotoniques, and Universitat Politecnica de Catalunya, Mediterranean Technology Park, 08860 Castelldefels (Barcelona), Spain

[2]Department of Physical Electronics, School of Electrical Engineering, Faculty of Engineering, Tel Aviv University, Tel Aviv 69978, Israel

We report that *defocusing* cubic media with spatially inhomogeneous nonlinearity, whose strength increases rapidly enough toward the periphery, can support stable *bright* localized modes. Such nonlinearity landscapes give rise to a variety of stable solitons in all three dimensions, including 1D fundamental and multihump states, 2D vortex solitons with arbitrarily high topological charges, and fundamental solitons in 3D. Solitons maintain their coherence in the state of motion, oscillating in the nonlinear potential as robust quasi-particles and colliding elastically. In addition to numerically found soliton families, particular solutions are found in an exact analytical form, and accurate approximations are developed for the entire families, including moving solitons.

    A commonly adopted principle underlying the studies of self-sustained localized modes (bright solitons) in various physical settings is that they are supported either by the focusing nonlinearity [1], or, in the form of gap solitons, by the defocusing nonlinearity combined with periodic linear potentials [2]. The formation of bright solitons was reported also in more sophisticated systems, where the nonlinearity periodically changes its magnitude, and even the sign, along the evolution variable or in the transverse direction(s). One thus deals with the *nonlinearity management* if it oscillates between focusing and defocusing in the course of the evolution [3], while transversely modulated nonlinearity landscapes are known as *nonlinear lattices* [4]. The latter setting readily supports stable solitons in 1D [5], while it is much harder to employ it for the stabilization of 2D and 3D solitons [6]. Note also that a hole in a uniform defocusing background was used as a support for 1D and 2D solitons in Ref. [8], but in a combination with a linear trapping potential.

    Guiding bright solitons by pure defocusing nonlinearities, without the help of a linear potential, is commonly considered impossible. The primary objective of this Rapid Communication is to demonstrate that this *is nevertheless possible*, if the strength of the defocusing term is modulated in space, growing fast enough towards the periphery. The existence of bright solitons in this setting is a consequence of the fact that, in contrast to media with homogeneous nonlinearities, when the presence of decaying tails of the soliton places it into the semi-infinite spectral gap of the linearized system, where defocusing nonlinearities cannot support any self-localization, in our case the growth of the nonlinearity coefficient makes the underlying equations *non-linearizable* for the decaying tails. A similar argument explains the existence of embedded solitons inside the continuous spectrum in self-focusing media [7].

    We demonstrate that the spatially modulated defocusing nonlinearity supports stable bright solitons, both quiescent and coherently moving ones, in all three dimensions. Not only fundamental solitons, but also stable 1D multipoles and 2D vortex rings are obtained. The model is based on the nonlinear-Schrödinger/Gross-Pitaevskii equation for rescaled field amplitude $q$ in optical media of dimension $D=1,2$, or the wave function in a Bose-Einstein condensate (BEC) of any dimension:

$$i\partial q/\partial \xi = -(1/2)\nabla^2 q + \sigma(\mathbf{r})|q|^2 q. \tag{1}$$

Here $\xi$ is the propagation distance or time, $\mathbf{r}=(\eta,\zeta,\tau)$ is the set of transverse coordinates, $\nabla^2 = \partial^2/\partial\eta^2 + \partial^2/\partial\zeta^2 + \partial^2/\partial\tau^2$, and $\sigma(\mathbf{r})>0$ is the defocusing nonlinearity strength that varies in the radial direction. In optics, spatially inhomogeneous nonlinearities can be realized in various ways [4]. In particular, in photorefractive materials, such as $LiNbO_3$, nonuniform doping with Cu or Fe may considerably enhance the local nonlinearity [9]. In BEC spatially modulated nonlinearity landscapes can be created, via the Feshbach resonance (FR), by nonuniform external fields [10,11]. We here assume that the strength of nonlinearity grows with radius as

$$\sigma(\mathbf{r}) = (\sigma_0 + \sigma_2 r^2/2)\exp(\alpha r^2) \tag{2}$$

with $\sigma_0, \sigma_2 > 0$, and $\alpha > 0$ that may be fixed by scaling (we set $\alpha = 1/2$ below). If the nonlinearity is controlled by the FR, the divergence of the nonlinearity strength at infinity implies that the background value of the control field at $r=\infty$ corresponds to the exact resonance. In the optical realization with dopants that give rise to the two-photon resonance, the effective modulation of the nonlinearity may be achieved via the inhomogeneity of the resonance detuning, controlled by an external field, with the exact resonance occurring at $r=\infty$.

For any $D$, Eq. (1) with $\sigma(\mathbf{r})$ taken as per Eq. (2) admits particular analytical solutions for fundamental solitons:

$$q(r,\xi) = (\alpha^2/\sigma_2)^{1/2}\exp(ib\xi - \alpha r^2/2), \tag{3}$$

with $b = -(D\alpha/2 + \sigma_0\alpha^2/\sigma_2)$, where $b$ is the propagation constant. For $\sigma_2 = 0$, exact solutions for a vortex with topological charge $m=1$ in 2D, and a dipole soliton in 1D are available too:

$$q(r,\xi) = (2\sigma_0)^{-1/2}\alpha r \exp(ib\xi + i\phi - \alpha r^2/2), \tag{4}$$

with $b = -\alpha(1+D/2)$, where azimuthal coordinate $\phi$ is a part of the solution for $D=2$.

The analytical solutions correspond to the particular values of propagation constant $b$ [multiplying Eq. (1) by $q^*$ and integrating, one can prove that the solitons may exist only for $b<0$]. For *families* of fundamental solitons, a variational approximation (VA) can be developed in any $D$ by adopting the ansatz suggested by the exact solutions, $q = A\exp(ib\xi - \alpha r^2/2)$ (amplitude $A$ is a variational parameter). Using the Lagrangian of Eq. (1), the VA yields the norm of the fundamental solitons as a function of $b$, written here for at $\sigma_2 = 0$ and $\sigma_0 = 1$:

$$U \equiv \int |q(\mathbf{r})|^2 d\mathbf{r} = -(\pi/\alpha)^{D/2}(b + \alpha D/2). \tag{5}$$

The comparison with numerical results presented in Figs. 1(c) and 3(c) demonstrates that the variational dependences $U(b)$ are virtually indistinguishable, on the scale of the figures, from their numerical counterparts for all dimensions. For the soliton's width, defined as $W = 2U^{-1} \int r|q(\mathbf{r})|^2 d\mathbf{r}$, the VA gives $W_{1D} = W_{3D}/2 = 2/(\pi\alpha)^{1/2}$, $W_{2D} = (\pi/\alpha)^{1/2}$. Numerically found widths approach these values with the increase of $U$, see, e.g., Fig. 5(b).

The steep anti-Gaussian profile of the modulation of the defocusing nonlinearity postulated in Eq. (2) is not a necessary condition for the existence of solitons. In fact $\sigma(r) \sim r^{D+\varepsilon}$ with arbitrary $\varepsilon > 0$, where $D$ is the spatial dimension, is sufficient [14]. Furthermore, for the exponential profile $\sigma(\eta) = a + \sinh^2(\eta)$, with any $a < 1$, it is easy to find an exact 1D soliton solution $w = (1-a)^{-1/2}\text{sech}(\eta)$ with $b = -(1+a)/[2(1-a)]$, and for $\sigma(\eta) = \cosh^2(\eta)$ one can find the exact dipole solution $w = 3^{1/2}\sinh(\eta)\text{sech}^2(\eta)$ with $b = -5/2$. The system can be also made finite, thus presenting a nonlinear counterpart of quantum-dot potentials. An example is the 1D variant of Eq. (1) with $\sigma(\eta) = (1/4)(1+3\eta^2)^2(1-\eta^2)^{-3}$, defined at $\eta^2 < 1$, which gives rise to an exact ground-state mode, $q(\eta,\xi) = (1-\eta^2)^2 \exp(-9i\xi/4)$.

Here we report numerical results for the basic version of model (2) with $\sigma_2 = 0$ and $\sigma_0 \equiv 1$. Fundamental solitons are sought for as $q(\mathbf{r},\xi) = w(r)\exp(ib\xi)$. The solutions were found using the standard relaxation method that quickly converges to exact solitons for a properly selected initial guess. The stability of thus found solutions was investigated by numerical computation of eigenvalues for small perturbations (with the help of an ordinary eigenvalue solver), using the linearization of Eq. (1), and then verified through direct simulations of the perturbed evolution.

As said above, our main result is that, in contrast to the belief that the defocusing nonlinearity cannot give rise to bright solitons, the inhomogeneous defocusing medium does support families of stable localized modes. The tails of the solitons of all types decay at $r \to \infty$ super-exponentially, irrespective of the dimension: $w|_{\eta \to \pm\infty} \approx (\alpha r/2^{1/2})\exp(-\alpha r^2/2)$, which complies with exact solutions (4). Note that this asymptotic form does not contain the propagation constant $b$.

Examples of 1D solitons, with different numbers $k$ of zeros (nodes) in the $w(\eta)$ shape, are displayed in Figs. 1(a) and 1(b). The solitons' amplitude increases with $|b|$, and the numerical results show that their width, at first, rapidly decreases and then saturates at $|b| \simeq 20$ (as predicted by the VA). For all types of the solitons, their energy flow (norm) increases with $|b|$ [Fig. 1(c)]. The solitons of higher orders have smaller norms, which is natural, taking into account the fact that, in terms of the mean-field description, the fundamental solitons, representing the ground state of the system, must minimize the chemical potential, $-b$, for a given norm.

The 1D solitons are remarkably robust. The computation of the stability eigenvalues demonstrates that the modes with $k = 0, 1, 2$ are stable at least up to $b = -40$ [in particular, this fact implies the stability of exact solution (4); it was checked that exact solution (3) is stable as well]. Only the families with $k \geq 3$ feature instability domains alternating with stability areas. The structure of the instability and stability domains becomes more complex with the increase of $k$, see Fig. 1(d) for $k = 5$. We did not find any limit on the number of nodes possible in stable 1D solitons, hence even very complex structures (with $k \geq 10$) may be stable. Direct simulations of the evolution of perturbed solitons verify the predictions of the stability analysis: while stable solitons keep their shape over distances far exceeding $\xi = 10^3$, their unstable counterparts transform into irregularly breathing modes that remain tightly confined, see the top row in Fig. 2.

The physically relevant definition of solitons includes their ability to maintain the intrinsic coherence in the state of motion, and quasi-elastic collisions. Solitons may be set in motion multiplying them by $\exp(i\theta\eta)$, with phase tilt $\theta$. As a result, both 1D and 2D solitons start regular oscillations (see examples for 1D solitons with $k = 0, 1, 2$ in the bottom row of Fig. 2) – somewhat similar to matter-wave solitons in the cigar-shaped traps [12], with the difference that the nonlinearity is repulsive in the present setting, and the solitons oscil-late in the effective nonlinear potential. An equation of motion for vectorial coordinate $\mathbf{R}(\xi)$ of the soliton can be readily derived in the quasi-particle approximation:

$$d^2\mathbf{R}/d\xi^2 = -2\alpha(\alpha/\pi)^{D/2} U_D \exp(2\alpha R^2)\mathbf{R}, \tag{6}$$

$D=1,2$ (here $U_D$ is the soliton's norm). As follows from Eq. (6), the squared frequency of small-amplitude oscillations of the kicked soliton is $\omega_D^2 = 2\alpha(\alpha/\pi)^{D/2}U_D + (3\alpha/2)\theta^2$, which was found to be in a virtually exact agreement with results of numerical simulations. Further, we applied opposite kicks to two lobes of a 1D dipole, thus initiating oscillations and recurrent collisions of two solitons with opposite signs. It was found that the solitons keep bouncing from each other elastically. Assuming the instantaneous rebound, Eq. (6) predicts the frequency of the periodic collisions very accurately too. Thus, both 1D and 2D solitons are robust quasi-particle objects, that maintain their intrinsic coherence in the course of the motion and interact elastically.

The 2D version of the model gives rise to vortex solitons, $q(r,\xi) = w(r)\exp(im\phi + ib\xi)$, for all integer values of topological charge $m$, see Figs. 3(a) and 3(b). For the same reason as in 1D, the vortices with different $m$, while having completely different asymptotic forms at $r \to 0$, become identical at $r \to \infty$ (in contrast to vortex solitons in focusing media, that considerably broaden with the increase of the topological charge [13]). The increase of $|b|$ results in a gradual contraction of the vortex rings toward $r=0$ [Fig. 3(b)]. The energy flow (norm) carried by the 2D solitons at fixed $b$ decreases with the increase of $m$ [Fig. 3(c)], similar to the 1D case, cf. Fig. 1(c).

Another essential result is that, due to the defocusing character of the nonlinearity, azimuthal instabilities, that are fatal for vortex solitons in focusing media [13], are suppressed in our system. We have found that the solitons with $m=0$ and $m=1$ [including the 2D exact solution (4)] are completely stable, while the vortices with $m>1$ give rise to a complex structure of stability and instability domains. This structure can be produced upon substituting a perturbed solution, $q = [w(r) + u(r)\exp(in\phi + \delta\xi) + v^*(r)\exp(-in\phi + \delta^*\xi)]\exp(im\phi + ib\xi)$, with azimuthal perturbation index $n$, into Eq. (1), and solving the corresponding linear eigenvalue problem. The structure of the stability domains is displayed in Fig. 3(d) for vortices with $m=2$, that can be destroyed by perturbations with $n=2$ at certain values of $b$ (similarly, at $m>2$ the most destructive perturbations pertain to $n=m, m\pm 1$). Note that the stability and instability domains are equidistantly spaced in $b$. We stress that conspicuous stability regions have been found for all the considered values of $m$. An example of the stable evolution of a perturbed vortex ring, which keeps its structure over indefinitely long distances, is shown in Fig. 4(a). Unstable vortex solitons (with $m \geq 2$) tend to split into $m$ separate unitary vortices, that stay in a vicinity of the pivotal point, performing persistent rotation around it, which is a consequence of the conservation of the angular momentum. Examples for $m=2$ and 3 are displayed in Figs. 4(b) and 4(c).

The 3D model also supports bright solitons with rapidly vanishing tails (recall the 3D model makes sense for BEC, but not in optics, unlike its 1D and 2D counterparts). Examples of such spherically symmetric fundamental solitons are shown in Fig. 5(a). The norm of the 3D solitons increases almost linearly with $|b|$, in accordance with Eq. (5), while their width rapidly saturates to the aforementioned VA-predicted value, $W_{3D} = 4/(\pi\alpha)^{1/2}$ [Fig. 5(b)]. The 3D fundamental solitons are completely stable in their entire existence domain, as illustrated by Fig. 6.

Summarizing, it is found that, in contrast to the usual expectations, the defocusing nonlinearity, without any linear potential, may support families of stable bright solitons in all dimensions, provided that the nonlinearity strength increases rapidly enough from the center to the periphery. In addition to the fundamental solitons, we show that such media support a variety of stable higher-order modes, including 1D multipoles and 2D vortex rings with all values of the topological charge. If set in motion, the solitons move and interact as particles. The settings considered here may be implemented for matter waves in BEC and for light waves in optical materials.

# Figure captions

Figure 1. (Color online) Profiles of 1D solitons: (a) with $b=-10$ and different numbers of nodes; (b) dipole solitons with different values of $b$. This and other figures are displayed for $\alpha=0.5$ in Eq. (2), with red horseshoe-shaped curves showing the nonlinearity modulation profile. (c) $U$ vs. $b$ for 1D solitons with different numbers of nodes, $k$. For $k=0$, this dependence is indistinguishable from its variational counterpart (5) with $D=1$. Here and in Fig. 3(c), stable and unstable portions of the soliton families are shown by black and green curves, respectively. (d) Stability (white) and instability (shaded) domains in the $(\alpha,b)$ plane for 1D solitons with $k=5$. The fan-shaped structure here and in Fig. 3(d) below is a manifestation of the scaling invariance of Eq. (1).

Figure 2. (Color online) Top row: Contour plots of $|q(\eta,\xi)|$ demonstrating the stable propagation of the perturbed 1D soliton with $k=1$, $b=-10$ (left), instability of the one with $k=3$, $b=-10$ (center), and stability of the complex mode with $k=5$, $b=-13$ (right). Bottom row: Oscillations of 1D solitons with $k=0,1,2$, $b=-20$, after the application of phase tilt $\theta=1.5$.

Figure 3. (Color online) Profiles of 2D solitons: (a) for $b=-10$ and different vorticities $m$; (b) for $m=2$ and different values of $b$. (c) $U$ vs. $b$ for different $m$ [the curve for $m=0$ is indistinguishable from the variational result (5) with $D=2$]. (d) The lowest stability (white) and instability (shaded) domains in the $(\alpha,b)$ plane for vortex solitons with $m=2$.

Figure 4. (Color online) (a) Stable propagation of the perturbed vortex soliton with $m=2$, $b=-17$. (b) Splitting of the unstable double vortex $(m=2)$ with $b=-11$ into a steadily rotating pair of unitary vortices. (c) Splitting of the unstable vortex with $m=3$, $b=-9$ into a rotating set of three vortices.

Figure 5. (Color online) (a) Profiles of fundamental 3D solitons at $\alpha=0.5$. (b) The width of these solitons vs. the norm.

Figure 6. Isosurface plots drawn at the level of $0.1\max|q|$ at $\xi=0$ (left), $\xi=300$ (center), and $\xi=600$ (right), showing stable propagation of the perturbed 3D soliton with $b=-10$.

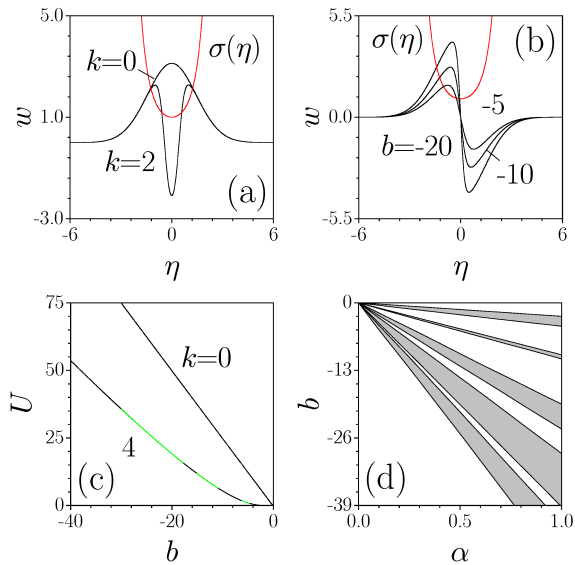

Figure 1. (Color online) Profiles of 1D solitons: (a) with $b=-10$ and different numbers of nodes; (b) dipole solitons with different values of $b$. This and other figures are displayed for $\alpha=0.5$ in Eq. (2), with red horseshoe-shaped curves showing the nonlinearity modulation profile. (c) $U$ vs. $b$ for 1D solitons with different numbers of nodes, $k$. For $k=0$, this dependence is indistinguishable from its variational counterpart (5) with $D=1$. Here and in Fig. 3(c), stable and unstable portions of the soliton families are shown by black and green curves, respectively. (d) Lowest stability (white) and instability (shaded) domains in the $(\alpha,b)$ plane for 1D solitons with $k=5$. The fan-shaped structure here and in Fig. 3(d) below is a manifestation of the scaling invariance of Eq. (1).

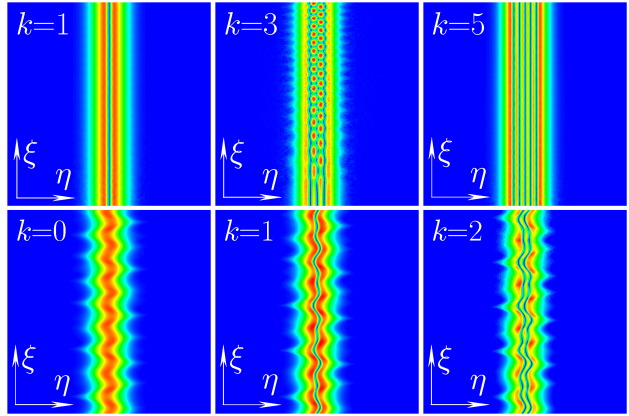

Figure 2.   (Color online) Top row: Contour plots of $|q(\eta,\xi)|$ demonstrating the stable propagation of the perturbed 1D soliton with $k=1$, $b=-10$ (left), instability of the one with $k=3$, $b=-10$ (center), and stability of the complex mode with $k=5$, $b=-13$ (right). Bottom row: Oscillations of 1D solitons with $k=0,1,2$, $b=-20$, after the application of momentum $\theta=1.5$.

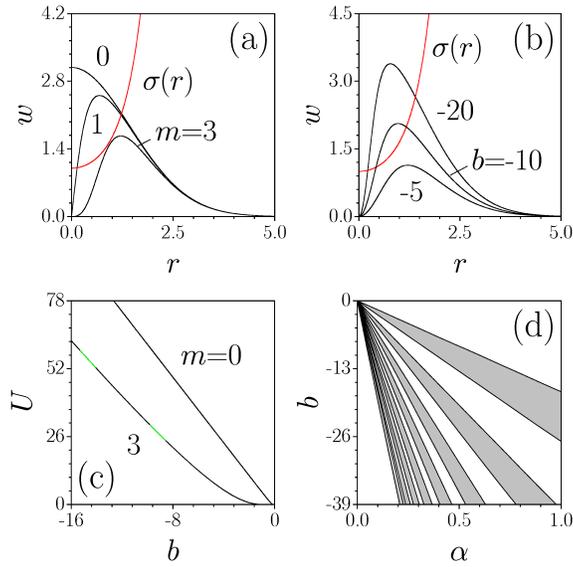

Figure 3.  (Color online) Profiles of 2D solitons: (a) for $b=-10$ and different vorticities $m$; (b) for $m=2$ and different values of $b$. (c) $U$ vs. $b$ for different $m$ [the curve for $m=0$ is indistinguishable from the variational result (5) with $D=2$]. (d) The lowest stability (white) and instability (shaded) domains in the $(\alpha,b)$ plane for vortex solitons with $m=2$.

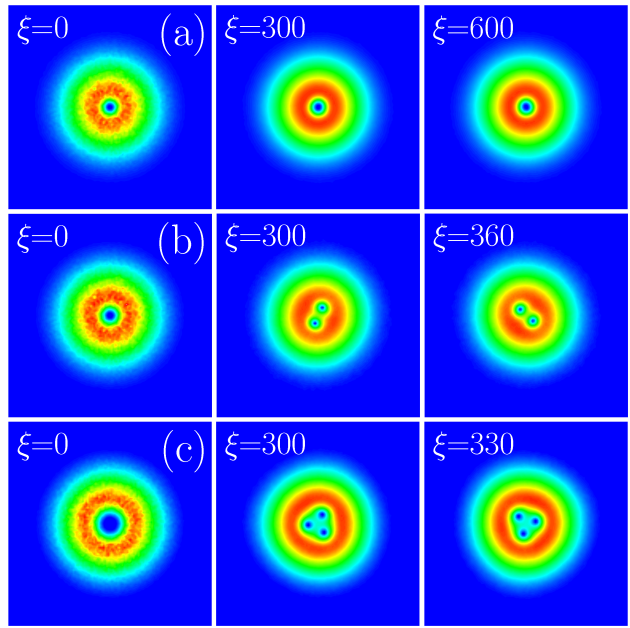

Figure 4. (Color online) (a) Stable propagation of the perturbed vortex soliton with $m=2$, $b=-17$. (b) Splitting of the unstable double vortex ($m=2$) with $b=-11$ into a steadily rotating pair of unitary vortices. (c) Splitting of the unstable vortex with $m=3$, $b=-9$ into a steadily rotating set of three vortices.

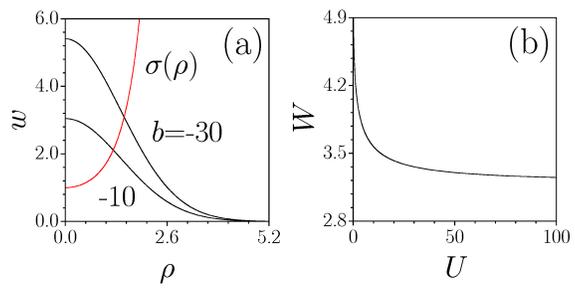

Figure 5. (Color online) (a) Profiles of fundamental 3D solitons at $\alpha = 0.5$. (b) The width of these solitons vs. the norm.

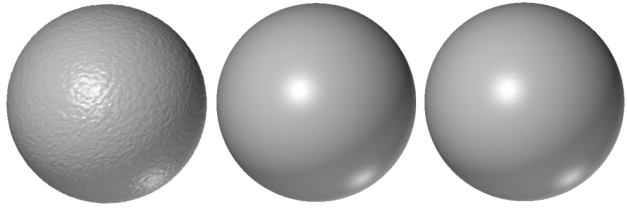

Figure 6.  Isosurface plots drawn at the level of $0.1\max|q|$ at $\xi=0$ (left), $\xi=300$ (center), and $\xi=600$ (right), showing stable propagation of the perturbed 3D soliton with $b=-10$.